\documentclass[10pt]{article}
\usepackage{algorithm}
\usepackage{algpseudocode}

\usepackage{amsmath}
\usepackage{fancyhdr}
\usepackage{makeidx}
\usepackage{amsthm}
\usepackage{amssymb}
\usepackage{amsfonts}
\usepackage{revsymb}
\usepackage{graphicx}
\usepackage{color}
\usepackage{ifthen}

\newtheorem{lemma}{Lemma}[section]
\newtheorem{theorem}[lemma]{Theorem}

\newcommand{\ket}[1]{{|#1\rangle}}
\newcommand{\bra}[1]{{\langle#1|}}

\newcommand{\oper}[1]{\hat{#1}}
\newcommand{\op}[1]{\hat{#1}}

\newcommand{\id}{{\oper{\openone}}}

\newcommand{\set}[1]{{\mathbf #1}}

\newcommand{\card}[1]{\left|#1\right|}             

\renewcommand{\algorithmicrequire}{\textbf{Requires:}}
\newcommand{\Input}{\item[\textbf{Input:}]}
\newcommand{\Participants}{\item[\textbf{Participants:}]}
\newcommand{\Goal}{\item[\textbf{Goal:}]}

\floatname{algorithm}{Protocol}

\begin{document}
\title{Anonymous Transmission of Quantum Information}
\author{Jan Bouda${}^{1,2}$\thanks{bouda@fi.muni.cz} and Josef {\v S}projcar${}^1$\thanks{xsprojc@fi.muni.cz}\\ \\
${}^1$Faculty of Informatics, Masaryk University\\ Botanick{\'a} 68a, 602 00 Brno, Czech Republic\\
Fax:+420-549 49 1820\\ \\
${}^2$ARC Seibersdorf research GmbH\\ Donau-City stra{\ss}e 1/5 og, 1220 Wien, Austria}
\maketitle
\begin{abstract}
We propose a protocol for anonymous distribution of quantum information that can be used to implement either channel with anonymous sender or channel with anonymous receiver. Our protocol achieves anonymity and message secrecy with unconditional security. It uses classical anonymous transfer. It tolerates disruption of the protocol, but the number of disrupters must be limited by the quantum Gilbert-Varshamov bound. This bound can be exceeded provided a specific entanglement distillation procedure will be used. A different version of the protocol tolerates any number of disrupters, but is secure only when receiver does not actively cooperate with other corrupted participants.
\end{abstract}
\newpage

\section{Introduction}

The amount of data transmitted via Internet or computer networks in general grew rapidly during last decades and it is expected that even more data influencing everyday life and economy will flow this way. On the other hand, computing and storage power of contemporary computers and improvement of algorithms allow automatic processing of huge amount of data and thus also filtering and tracing particular messages. These can be selected according to sender, receiver or even content. While the content of a message can be secured using various methods of encryption, sometimes already the existence of the message as well as the identity of its sender and recipient are sensitive information. By following recipients of your messages one can determine a lot about your personal or professional life.

As a reaction on such an abuse of computer networks a number of methods assuring user (i.e. sender and/or receiver) anonymity were proposed. These methods include both computationally \cite{Chaum-UntraceableElectronicMail-1981} and unconditionally secure solutions \cite{Chaum-DiningCryptographersProblem:Unconditional-1988,Bos.Boer-DetectionofDisrupters-1990,Waidner.Pfitzmann-UnconditionalSenderand-1990,Waidner.Pfitzmann-DiningCryptographersin-1989}. While the latter are usually less efficient, they are the only way to ensure provable anonymity of sender and/or receiver. This establishes them as an important part of modern cryptography and justifies the attention they have received during last twenty years. It is slightly surprising that applications of these primitives go far beyond preserving anonymity of communicating parties. Anonymous transfer protocols are known to have wide applications in design of secure ballot elections \cite{Pfitzmann.Waidner-UnconditionallyUntraceableand-1992}, Byzantine agreement \cite{Pfitzmann.Waidner-UnconditionalByzantineAgreement-1992} or key distribution protocols \cite{Alpern.Schneider-KeyExchangeusing-1983} and are the key ingredient of digital pseudosignatures \cite{Chaum.Roijakkers-Unconditionally-SecureDigitalSignatures-1991}.

In this paper we propose a protocol realizing anonymous transfer of quantum information, which has equivalent importance for future quantum networks as preserving anonymity in classical networks. Using anonymous transfer of classical information as a primitive\footnote{It means we use it as a blackbox. Note that composite protocol can be only as secure as its weakest part so the security of the primitive may limit security of the composite protocol.} it assures anonymity, secrecy of the message and detection of disrupters (also tracelessness, see below).

Before continuing with explanation of our approach we would like to stress differences to the paper \cite{Christandl.Wehner-QuantumAnonymousTransmissions-2005}, where authors independently proposed a number of ideas and techniques contained also in our paper. Main results of the paper \cite{Christandl.Wehner-QuantumAnonymousTransmissions-2005} are two protocols, one for anonymous transfer of classical information and one for anonymous transfer of quantum information. The main motivation was to design traceless\footnote{Protocol is traceless if it remains secure (anonymous) even after all participants publish all random bits they used.} protocols, while sacrificing security of the message and detection of disrupters. Also, authors did not design the protocols from scratch neither based them on some well-established cryptographic primitives. The initial assumption of these protocols is that participants share securely created and distributed generalized GHZ state. Distribution of this state is an important part of our protocol, where one can achieve security stronger that the one proposed in \cite{Christandl.Wehner-QuantumAnonymousTransmissions-2005}.

The paper \cite{Christandl.Wehner-QuantumAnonymousTransmissions-2005} demonstrates that quantum information processing might offer resources allowing anonymous transfer of classical information with security features not achievable by classical cryptography. Recently, there were proposed also other quantum techniques \cite{Vaccaro.Spring.ea-Quantumprotocolsanonymous-2005,Hillery.Ziman.ea-Towardsquantum-basedprivacy-2005}, which partly establish anonymous transfer protocol for classical information. While these solutions achieve no security improvements in contrast to classical protocols, they propose alternative ways of founding anonymous transfer on principles of quantum mechanics.

\section{Terminology and definitions}

\subsection{General approach to multi-party cryptographic protocols}
Before introducing anonymous transfer problem we will briefly recapitulate general approach to secure multi-party computation, which is a nice and uniform way of modeling cryptographic protocols, especially adversaries. In what follows we will denote $\set{P}$ the set of all participants of the protocol.
The main goal of multi-party cryptographic protocols is to ensure that honest participants achieve the goal specified in the definition of the problem regardless of actions of dishonest participants.

In order to model that all malicious participants fully cooperate, we interpret the cheating in the way that the (active) adversary is a participant not included in the set $\set{P}$ who takes full control of some subset $\set{C}\subset\set{P}$ of participants\footnote{When adversary corrupts some set of participants, it models the practical situation when all members of this set of participants are dishonest and are fully cooperating in order to compromise security of the protocol.} of the protocol. He is completely controlling participants in the set $\set{C}$, especially he has access to their internal memory, can read their inputs and controls their outputs. This approach was introduced in context of composable security, see e.g. \cite{Canetti-unifi_frame_analy:2001} and references included therein.

In this approach we can distinguish active and passive adversaries. Passive adversary controls the set of corrupted participants only in a limited way -- he can read all their inputs, outputs and internal information, but he cannot change behavior of corrupted participants. In this sense the protocol \cite{Christandl.Wehner-QuantumAnonymousTransmissions-2005} is secure against passive adversary. Active adversary has all abilities as the passive one, and, in addition, he can change behavior of corrupted participants. Our protocol is safe against active adversary.

There also exists an alternative approach to model cheating in protocols. This approach is more traditional and sometimes allows to explain more easily cheating objectives of groups of participants, but it is less formal. Cheating objectives are usually specified by description of potential cooperating sets (collusions) of malicious participants together with goals they are willing cooperate on.

In our paper we will use both approaches, we use the single adversary approach mainly to explain generality of our adversary model while the collusion-based approach is used in concrete situations to explain cheating objectives of single participants of collusions. The switching between both approaches is easy and natural.

Finally, we would like to mention that in the setting of complex cryptographic protocols (e.g. anonymous channels) we are not interested in adversary able to tamper with bipartite communication channels, namely we require that he is not able to modify bipartite communication. It makes no harm if he can read the communication, however, in quantum case authentication of a particular channel implies also its encryption \cite{Barnum+Crepeau...-Authenticatio_of_q_mes:2002}.

\subsection{Definition of anonymous message transfer}

The term anonymous transfer includes a whole range of different subproblems. In order to be consistent we adopted the terminology introduced in \cite{Pfitzman.Hansen-AnonymityUnlinkabilityUnobservability-}. In this subsection we will briefly review part of this terminology necessary for our paper.

{\bf Anonymity} of an object is the state of being not identifiable within a set of subjects known as the {\bf anonymity set}. Anonymity set usually consists of parties able to perform a particular action we are interested in. In our case it means that the sender (receiver) is not identifiable within a particular set of potential senders (receivers).

The concept of unconditionally secure anonymity was introduced by Chaum \cite{Chaum-DiningCryptographersProblem:Unconditional-1988} as the problem of dining cryptographers and all solutions based on the technique introduced in this paper are denoted as DC-nets.
There is a number of different anonymous channels, we will consider primarily three basic versions.

{\bf Anonymous broadcast} was the original setting introduced in \cite{Chaum-DiningCryptographersProblem:Unconditional-1988}. There are $n$ participants in this problem -- $n$ potential senders $\set{S}=\{P_i\}_{i=1}^n$ and one real anonymous sender $S\in\set{S}$. $S$ has a message $msg$ and wants to broadcast it in the way that
\begin{itemize}
\item Everyone receives $msg$.
\item When adversary corrupts some set of participants $\set{C}\subset\set{S}$, then the anonymity set\footnote{This is quite strict, often the anonymity set might be smaller, however, this is precisely what is achievable classically and what our protocol achieves.} of the sender is $\set{S}\smallsetminus\set{C}$.
\item No one (except sender) can {\bf disrupt}\footnote{I.e. to behave in the way that he prevents correct execution of the protocol.} the protocol without being detected.
\end{itemize}

In {\bf message transfer with anonymous sender (MTAS)} there are $n$ participants -- $(n-1)$ potential senders $\set{S}=\{P_i\}_{i=1}^{n-1}$, one real anonymous sender $S\in\set{S}$ and one publicly known receiver $R\in\set{P}=\set{S}\cup\{R\}$. $S$ has a message $msg$ and he wants to transmit it to $R$ in the way that
\begin{itemize}
\item $R$ receives $msg$.
\item No one except $S$ and $R$ learns the message.
\item When adversary corrupts some set of participants $\set{C}\subset\set{P}$, then the anonymity set of the sender is $\set{S}\smallsetminus\set{C}$.
\item No one (except $S$ and $R$) can disrupt the protocol without being detected.
\end{itemize}

{\bf Message transfer with anonymous receiver (MTAR)} is close to the previous problem. There are $n$ participants -- $(n\!-\!1)$ potential receivers $\set{R}\!=\!\{P_i\}_{i=1}^{n-1}$, one anonymous receiver $R\!\in\!\set{R}$ and one publicly known sender $S\!\in\!\set{P}\!=\!\set{R}\!\cup\!\{S\}$. $S$ has a message $msg$ and he wants to transmit it to $R$ in the way that
\begin{itemize}
\item $R$ receives $msg$.
\item No one except $S$ and $R$ learns the message.
\item When adversary corrupts some set of participants $\set{C}\subset\set{P}$, then the anonymity set of the receiver is $\set{R}\smallsetminus\set{C}$.
\item No one (except $S$ and $R$) can disrupt the protocol without being detected.
\end{itemize}
All these primitives can be classically realized with unconditional security against active adversary using solutions based on the dining cryptographer nets.

In this paper we propose protocol for MTAS of quantum information and in conclusion we show how to easily modify it to realize MTAR of quantum information. Before proceeding to the design of the protocol we will briefly discuss cheating objectives of participants of the protocol in the case of MTAS.

The sender $S$ is the completely trusted participant in this problem - he has no cheating objectives. He knows all information, he has no reason to disrupt the protocol and also he has no reason to wrongfully accuse some other participant of disruption -- this will only decrease his anonymity set\footnote{Important is that once the sender is (possibly) changed, all participants previously excluded (because of disruption) may join the protocol again. Otherwise sender may accuse honest participant of disruption to reduce anonymity set of the next sender.}.

$R$ is not interested in disruption of the protocol with one exception - he might be interested to disrupt the protocol provided that he can accuse other participant of disruption and exclude him from the protocol. $R$ also does not have to cheat in order to learn the message - he will learn it anyway. His main goal remains to learn the identity of the sender $S$.

Other participants - potential senders - might be interested in a number of cheating objectives - they want to learn the identity of $S$, they want to learn the message $msg$ and they want to disrupt the protocol.

\section{Design of the Protocol and its Security}
\label{sec:Qanon_protocol}
Let us first discuss reasonability of the initial assumption of \cite{Christandl.Wehner-QuantumAnonymousTransmissions-2005}. In the case we do not consider disrupters, this state can be easily created - it suffices that one randomly chosen participant creates and distributes it. He is trusted to do because by distributing different state he does not compromise anonymity of the sender and we suppose that participants are not interested in disruption of the protocol. Once we consider possible disruption, this assumption starts to be problematic. In case of anonymous channel for classical information it is not achievable, since existence of this state will imply unbiased multipartite coin tossing, what is not possible \cite{Ambainis+Buhrman...-Multi_quant_coin:2003} in presence of dishonest participants. In case of anonymous transfer of quantum information the situation seems to be similar, but the anonymous sender can act through classical anonymous channel to deliver trusted information to other participants. It might be interesting to investigate whether this changes the situation, although it does not seem to be probable.

The first idea of our protocol, proposed also by \cite{Christandl.Wehner-QuantumAnonymousTransmissions-2005}, is that $S$ can easily transfer any quantum state anonymously to $R$ provided that $S$ and $R$ share an EPR pair, but $R$ does not know who is controlling the other system. To transfer a state $\op{\rho}$ $S$ simply teleports it using the EPR pair to $R$ and sends him the result of the measurement using anonymous broadcast, as described in Protocol \ref{alg:quantum_MTAS}. To implement an anonymous transmission of quantum information it remains to design a protocol achieving anonymous sharing of an EPR pair.

\begin{algorithm}
\caption{MTAS of quantum information}
\label{alg:quantum_MTAS}
\begin{algorithmic}[1]
\Participants Potential senders $\set{S}$, publicly known receiver $R$, unknown sender $S\in\set{S}$
\Input quantum system $M$
\Goal $S$ wants to send the state of $M$ to $R$ in the way that his identity remains secret.
\Require anonymous broadcast of classical information, EPR pair $\ket{\Phi^+}_{SR}$ shared between $S$ and $R$ such that the identity of $S$ is unknown (even to $R$).
\Statex
\State Sender $S$ performs Bell measurement on the systems $M$ and $S$ and sends the outcome of the measurement to $R$ using anonymous message broadcast.
\State $R$ uses this information to reconstruct the original state.
\end{algorithmic}
\end{algorithm}

In order to explain the design and security proof of our protocol in a transparent way we will introduce first highly simplified Protocol \ref{alg:anonymous_noisy_EPR_generation_ndd} capturing the basic principle of the Protocol \ref{alg:anonymous_EPR_generation}. Protocol \ref{alg:anonymous_noisy_EPR_generation_ndd} in fact solves anonymous transfer of quantum information when we do not consider disrupters, but instead of assuming initially shared generalized GHZ state it uses existing classical solution of anonymous channel.

\begin{algorithm}
\caption{Anonymous noisy EPR generation without disruption detection}
\label{alg:anonymous_noisy_EPR_generation_ndd}
\begin{algorithmic}[1]
\Participants Potential senders $\set{S}=\{P_i\}_{i=1}^{n-1}$, publicly known receiver $R$, unknown sender $S\in\set{S}$
\Goal $S$ wants to generate EPR pair(s) shared with $R$ while staying anonymous.
\Require MTAR of classical information,  bipartite authenticated quantum channel between each pair of participants
\Statex
\State Each $P_i\in\set{S}$ creates $n$ qubit quantum system in the state $\ket{\Phi}=\frac{1}{\sqrt{2}}(\ket{0}^{\otimes n}+\ket{1}^{\otimes n})$, keeps one subsystem and sends to each participant (including $R$) one of the remaining subsystems.
\State Each potential sender $P_i$ (except $S$) measures all systems he is possessing in the basis $\{\ket{+}, \ket{-}\}$ and records the outcomes $o_{i,j},j=1\dots n$ (each participant received $n$ quantum systems). He sends these results to $S$ using MTAR.
\State $S$ sends himself a dummy message (e.g. a random bit) using MTAR.
\State $S=P_{j'}$ repairs the state of each system he shares with $R$ to the EPR pair $\ket{\Phi^+}$. To repair the $i$--th system he applies $\sigma_z$ to his system if $\oplus_{j=1}^{n-1} o_{j,i} = 1$ (we formally set $o_{j',i}=0$).
\end{algorithmic}
\end{algorithm}

The basic method of the generation of an EPR pair in Protocol \ref{alg:anonymous_noisy_EPR_generation_ndd} is based on a simple observation. If we take an $n$--qubit generalized GHZ state $\ket{\Phi}=\frac{1}{\sqrt{2}}(\ket{0}^{\otimes n}+\ket{1}^{\otimes n})$ and measure arbitrary $(n-2)$ qubits independently in the dual basis $\left\{\ket{+}=\frac{1}{\sqrt{2}}(\ket{0}+\ket{1}), \ket{-}=\frac{1}{\sqrt{2}}(\ket{0}-\ket{1})\right\}$, then the state of two remaining qubits collapses, depending on results of the performed measurements, either to $\ket{\Phi^+}=\frac{1}{\sqrt{2}}(\ket{00}+\ket{11})$ (if the number of outcomes $\ket{-}$ was even) or to $\ket{\Phi^-}=\frac{1}{\sqrt{2}}(\ket{00}-\ket{11})$ (if the number of outcomes $\ket{-}$ was odd). Sender receives outcomes of all measurements and hence he can easily repair the state of the system $SR$ to $\ket{\phi^+}$ by applying $\sigma_z$ to his system when necessary (he counts the number of outcomes $\ket{-}$).

Let us suppose that measurement outcomes are publicly broadcasted instead of being sent via MTAR. In this case also the sender is forced to broadcast publicly outcome of his measurement. Since the distributor of the state $\ket{\Phi}$ is in principle interested in compromising anonymity of the sender, he may create and distribute system of $n$ qubits, where each of the qubits will be randomly in a state $\ket{+}$ or $\ket{-}$. Since sender does not perform measurement, he has chance $1/2$ to be caught not measuring the system and his identity will be revealed. It is not sufficient to send the measurement outcomes securely to receiver, because he can cheat as well. The key obstacle is that it is insecure to suppose that the original system is in the state $\ket{\Phi}$ if it is created by someone else than the sender. Therefore we use MTAR to send outcomes to $S$.

The fact that quantum system in state $\ket{\Phi}$ is created and distributed by each of the participants will be used in Protocol \ref{alg:anonymous_EPR_generation} to lay traps to detect disrupters while preserving anonymity of the sender.

\begin{theorem}
\label{the:anonymous_noisy_EPR_generation_ndd}
Protocol \ref{alg:anonymous_noisy_EPR_generation_ndd} preserves anonymity of the sender and when participants do not disrupt it, it creates $n-1$ EPR pairs shared between public receiver and anonymous sender.
\end{theorem}
\begin{proof}
Behavior of the sender differs only in local actions he performs on his quantum systems, what cannot be detected, and in the classical bit he sends in Step $4\!\!:$, that is uncorrelated to the quantum system he received. However, he is the only receiver of this information and hence his behavior is indistinguishable from all other participants.

It is obvious that when other participants do not disrupt the protocol, it creates $n-1$ EPR pairs shared between public receiver and anonymous sender.
\end{proof}
The disadvantage of Protocol \ref{alg:anonymous_noisy_EPR_generation_ndd} is that it is vulnerable even to a small amount of noise, which can be introduced by a malicious participant. To prevent this we have to allow honest participant to detect possible malicious behavior of other players on the state he distributed. This means we should provide him a way how to verify that they indeed performed measurement in dual basis and submitted correct outcome to the sender. In order to do this each player in each round of Protocol \ref{alg:anonymous_EPR_generation} randomly decides whether he will operate in an actual or in a trap mode. In the trap mode he sends his trap states to the sender who later verifies whether there was any disagreement with reported outcomes of particular measurements. Note that this disagreement can be caused by both disruption and malicious behavior of the trap layer, but this does not compromise security of the protocol.

Important is that if an honest trap state was distributed, malicious participants must either measure it in the dual basis and submit correct outcome or there is a probability that they will be detected, which is proportional to the amount of noise (strength of disruption) they introduce. This is formulated in Lemma \ref{lem:qanon:disrupt_or_detect}, which is proven in Appendix \ref{sec:Proof-Lemma}.

\begin{lemma}
\label{lem:qanon:disrupt_or_detect}
Let us suppose that quantum systems $P_1,\dots,P_n$ (controlled by a set of participants $\set{P}$) are either in the generalized GHZ state $1/\sqrt{2}(\ket{0^n}+\ket{1^n})$ or in one of the states $\{\ket{+},\ket{-}\}^{\otimes n}$. Let us further suppose that there is some collusion $\set{C}\subset\set{P}$ of corrupted participants controlling (WLOG) quantum systems $P_1,\dots,P_m$, $m<n$. Let us suppose that they perform some quantum operation that distinguishes with good probability states $\{\ket{+},\ket{-}\}^{\otimes m}$ from each other. Provided this operation gave outcome $\ket{\phi}\in\{\ket{+},\ket{-}\}^{\otimes m}$, the state of unmeasured systems $P_{m+1},\dots,P_n$ is close to
\begin{equation}
Tr_{\set{C}}\left(\frac{\left[\ket{\phi}_{\set{C}}\bra{\phi}\otimes\id_{\set{P}\smallsetminus\set{C}}\right]\ket{\Phi}}{\bra{\Phi}\left[\ket{\phi}_{\set{C}}\bra{\phi}\otimes\id_{\set{P}\smallsetminus\set{C}}\right]\ket{\Phi}}\right).
\end{equation}
\end{lemma}

When disagreement is detected, protocol is restarted and potential sender will simply ignore the participant who was detected to disrupt his state and therefore this participant has no chance to introduce errors. The list of participants each potential sender $P_i$ distributes the generalized GHZ state to is the set $\set{P}_i$. This is analogical to removing edge from the key sharing graph in \cite{Chaum-DiningCryptographersProblem:Unconditional-1988}.

\begin{theorem}
\label{the:anon_EPR_generation}
Protocol \ref{alg:anonymous_EPR_generation} preserves anonymity of sender and it with high probability generates at least one (concrete number of EPR pairs depends on entanglement distillation settings) EPR pair shared between anonymous sender and public receiver for each honest potential sender.
\end{theorem}
\begin{algorithm}
\caption{Anonymous EPR generation}
\label{alg:anonymous_EPR_generation}
\begin{algorithmic}[1]
\Participants Potential senders $\set{S}=\{P_i\}_i$, publicly known receiver $R$, unknown sender $S\in\set{S}$
\Input $m$ -- the number of rounds that should be executed, $p$ - probability that potential sender decides to switch to the trap mode
\Goal $S$ wants to generate a number of EPR pairs shared with $R$ in the way that his identity remains anonymous.
\Require MTAR of classical information, anonymous broadcast, bipartite authenticated quantum channel between each pair of participants, public entanglement distillation protocol
\Statex
\State Set all sets $\set{P}_i=\set{P}$.
\Repeat \Comment{Repeat $\ m$ times}
\State Each potential sender $P_i$ (including $S$) decides whether he will operate in trap (probability $p$) or actual (probability $1-p$) mode and reports this to $S$ using MTAR.
\State Each $P_i$ operating in actual mode creates an $\card{\set{P}_i}$ qubit quantum system in the state $\ket{\Phi}=\frac{1}{\sqrt{2}}(\ket{0}^{\otimes \card{\set{P}_i}}+\ket{1}^{\otimes \card{\set{P}_i}})$, keeps one subsystem and sends to each participant in $\set{P}_i$ (including $R$) one of the remaining subsystems.
\State Each $P_i$ operating in trap mode sends to each participant in $\set{P}_i$ quantum system randomly in state $\ket{+}$ or $\ket{-}$ and remembers states he sent.
\State Each potential sender $P_i$ (except $S$) measures all systems he is possessing in the basis $\{\ket{+}, \ket{-}\}$ and records outcomes $o_{i,j},j=1\dots n-1$ (each participant received $n-1$ quantum systems). First index refers to participant performing the measurement, second index identifies creator (distributor) of the system. He sends these results to $S$ using MTAR.
\State Sender measures all systems which were reported in Step $3\!:$ to be trap states.
\State Each potential sender operating in the trap mode sends to $S$ using MTAR the list of trap states he distributed.
\State $S$ compares each received list with outcomes of measurements he obtained from participants and publicly (using anonymous broadcast) announces any disagreement including disagreement in his measurement.
\State If there was any disagreement, corresponding sets $\set{P}_i$ are modified and protocol is restarted: If there was a trap system distributed by participant $P_i$ and incompatible measurement outcome received from participant $P_j$, then $P_j$ will be removed from $P_i$'s cooperation set, i.e. $\set{P_i}\gets\set{P_i}\smallsetminus\{P_j\}$.
\State $S=P_{j'}$ repairs the state of each system he shares with $R$ and which was not reported to be the trap state to the EPR pair $\ket{\Phi^+}$ (in case the system was not disrupted, otherwise the state is random), i.e. to repair the $i$--th system he applies $\sigma_z$ to his system if $\oplus_{j=1}^{n-1} o_{j,i} = 1$ (we formally set $o_{j',i}=0$).
\State Both $S$ and $R$ remember who created each particular EPR pair (more precisely - the original state $\ket{\Phi}$). Later, in Step $14\!\!:$, they perform independently a test for purity of EPR pairs created by each participant.
\Until{executed $m$ times}
\State $S$ and $R$ perform entanglement distillation with one-way communication independently on each set of all (possibly noisy) EPR pairs created by particular participant. They succeed with high probability to distill EPR states created by each honest participant. Classical communication from $R$ to $S$ is realized using MTAR.
\State Create anonymously shared EPR pair(s), see discussion following after Theorem \ref{the:anon_EPR_generation}.
\end{algorithmic}
\end{algorithm}
The last step of Protocol \ref{alg:anonymous_EPR_generation} allows three possible solutions with different security implications. In Step $14$ sender learns which of sets was distillable \cite{Bennett+Brassard...-Purif_noisy_entan:1995} and which was not. Important is that receiver does not learn this and that is why entanglement distillation with one-way communication must be used. The number of EPR pairs dishonest participants must disrupt grows with the number of generated EPR pairs. On the other hand, the probability of detection of at least one disrupter grows exponentially with the number of EPR pairs malicious participants disrupts and therefore we can always guarantee reasonable amount of noise (that allows distillation) in the set of states which was generated by an honest participant.

We can modify the protocol in the way that it is not necessary to restart it when disrupter is detected. The only difference is that disrupters are able to introduce more noise than in the original protocol, each eavesdropper must be considered independently and therefore possible undetected noise in each particular ensemble will be multiplied by the number of disrupters. This implies that the number $m$ of rounds must be increased appropriately. On the other hand, the fact that we do not have to restart the protocol increases efficiency greatly, since disrupters might behave in the way that they disrupt at the end of the protocol to increase time and communication complexity.

The key problem is the Step $15\!:$ since sender and receiver must establish a number of pure EPR pairs, but receiver should not be able to learn which set of original states was not distillable. In case receiver can do so, he can in cooperation with other dishonest participants compromise anonymity of the sender. It is sufficient that each dishonest participant introduces in the states he generates noise affecting only one honest participant, e.g. he is sending him random particle and otherwise he behaves correctly. In case such an ensemble of states is found not distillable in Step $14$, it means that the only disrupted participant was the sender and his identity is revealed.

In order to prevent this receiver must not learn which ensemble was noisy and which was not. Possible solution to this is to take all states obtained in Step $14$ regardless whether they emerged from successful or insuccessful distillation and distill them together using some distillation procedure with one-way communication. The number of errors is the number of system generated by disrupters. In case we consider that we extracted one EPR pair for each participant, the number of disrupting participants protocol can tolerate is bounded by the quantum Gilbert-Varshamov bound which establishes the bound for errors distillable with one-way communication \cite{Ambainis.Gottesman-MinimumDistanceProblem-2003}. A subject of future research will be to design specific entanglement distillation procedure that uses the sender's knowledge which states are pure and which are noisy. This may allow to exceed the Gilbert-Varshamov bound.

Other possibility is to simply inform receiver in which cases the distillation succeeded. In this case protocol works even in case there are only two nondisrupting potential senders, but it is secure only provided that receiver does not cooperate in a malicious way with potential senders.

\section{Conclusion}
Protocol \ref{alg:anonymous_EPR_generation} is quite difficult to implement experimentally, however, for anonymous transfer without disruption detection it is sufficient to use simplified version of Protocol \ref{alg:anonymous_noisy_EPR_generation_ndd}, especially it is sufficient that only one randomly chosen participant distributes the generalized GHZ state. Five participant version of this protocol is (up to classical communication) exactly the experiment \cite{Zhao.Chen.ea-Experimentaldemonstrationof-2004}, although the motivation of this experiment was different. Therefore the experiment \cite{Zhao.Chen.ea-Experimentaldemonstrationof-2004} can be considered to be the first experimental realization of simplified version of our anonymous transfer protocol.

It is possible to adapt Protocol \ref{alg:anonymous_EPR_generation} to tolerate certain amount of noise. All what has to be done is that small fraction of incorrect answers won't be considered to be disruption and parameters $p$ and $m$ (optionally also distillation parameters) will be adjusted to increase probability of disrupter detection and preserve the level of noise.

Due to no-cloning theorem and properties of authentication of quantum information \cite{Barnum+Crepeau...-Authenticatio_of_q_mes:2002} it is impossible to simply modify the digital pseudosignatures proposed in \cite{Chaum.Roijakkers-Unconditionally-SecureDigitalSignatures-1991} to work with quantum information. There are two main reasons - from the no-cloning theorem we obtain that we must have a number of identical copies of quantum state if we want to apply digital pseudosignatures. Even more fundamental problem is the result of \cite{Barnum+Crepeau...-Authenticatio_of_q_mes:2002} that any authentication also encrypts quantum information and therefore other participants have no chance to verify whether two different signatures correspond to the same message.

As a conclusion we remark that it is possible to modify this protocol to realize MTAR of quantum information. It suffices to invert the direction in which the message is being teleported at the end of the protocol. Our approach can be also modified to assure simultaneous anonymity of the sender and the receiver, but this beyond scope of this paper.

\section{Acknowledgements}
We thank J\"orn M\"uller-Quade, Andreas Poppe, Harald Weinfurther and Andreas Winter for useful and stimulating discussions. Support of the grant projects GA\v CR 201/04/1153 and GA\v CR 102/05/H050 is acknowledged. J.B. acknowledges support of the Hertha Firnberg ARC stipend program and grant project GA\v CR 201/06/P338.
\bibliographystyle{splncs}
\bibliography{../../../bibliografie/qcrypto}

\begin{thebibliography}{10}

\bibitem{Chaum-UntraceableElectronicMail-1981}
Chaum, D.:
\newblock Untraceable electronic mail.
\newblock Communications of the ACM \textbf{24}(2) (1981)  84--88

\bibitem{Chaum-DiningCryptographersProblem:Unconditional-1988}
Chaum, D.:
\newblock The dining cryptographers problem: Unconditional sender and recipient
  untraceability.
\newblock J. Cryptology \textbf{1}(1) (1988)  65--75

\bibitem{Bos.Boer-DetectionofDisrupters-1990}
Bos, J., den Boer, B.:
\newblock Detection of disrupters in the {D}{C} protocol.
\newblock In: EUROCRYPT'89. (1990)  320--327 LNCS 434.

\bibitem{Waidner.Pfitzmann-UnconditionalSenderand-1990}
Waidner, M., Pfitzmann, B.:
\newblock Unconditional sender and recipient untraceability in spite of active
  attacks.
\newblock LNCS \textbf{434} (1990)  302--362

\bibitem{Waidner.Pfitzmann-DiningCryptographersin-1989}
Waidner, M., Pfitzmann, B.:
\newblock The dining cryptographers in the disco: Unconditional sender and
  recipient untraceability with computationally secure serviceability.
\newblock In: EUROCRYPT'89. (1989) extended abstract.

\bibitem{Pfitzmann.Waidner-UnconditionallyUntraceableand-1992}
Pfitzmann, B., Waidner, M.:
\newblock Unconditionally untraceable and fault-tolerant broadcast and secret
  ballot election.
\newblock Technical Report 3/92, Hildesheimer Informatik-Berichte (1992)

\bibitem{Pfitzmann.Waidner-UnconditionalByzantineAgreement-1992}
Pfitzmann, B., Waidner, M.:
\newblock Unconditional byzantine agreement for any number of faulty
  processors.
\newblock In: Symposium on Theoretical Aspects of Computer Science. (1992)
  339--350

\bibitem{Alpern.Schneider-KeyExchangeusing-1983}
Alpern, Schneider:
\newblock Key exchange using 'keyless cryptography'.
\newblock Information processing letters \textbf{16} (1983)  79--81

\bibitem{Chaum.Roijakkers-Unconditionally-SecureDigitalSignatures-1991}
Chaum, D., Roijakkers, S.:
\newblock Unconditionally-secure digital signatures.
\newblock In: CRYPTO'90. (1991)  206--214 LNCS 537.

\bibitem{Christandl.Wehner-QuantumAnonymousTransmissions-2005}
Christandl, M., Wehner, S.:
\newblock Quantum anonymous transmissions.
\newblock In: 11th ASIACRYPT. LNCS 3788 (2005)  217--235 quant-ph/0409201.

\bibitem{Vaccaro.Spring.ea-Quantumprotocolsanonymous-2005}
Vaccaro, J.A., Spring, J., Chefles, A.:
\newblock Quantum protocols for anonymous voting and surveying.
\newblock quant-ph/0504161 (2005)

\bibitem{Hillery.Ziman.ea-Towardsquantum-basedprivacy-2005}
Hillery, M., Ziman, M., Buzek, V., Bielikova, M.:
\newblock Towards quantum-based privacy and voting.
\newblock Physics Letters A (349) (2006)  75--81 quant-ph/0505041.

\bibitem{Canetti-unifi_frame_analy:2001}
Canetti, R.:
\newblock A unified framework for analyzing security of protocols.
\newblock Electronic Colloquium on Computational Complexity \textbf{Report No.
  16} (2001)

\bibitem{Pfitzman.Hansen-AnonymityUnlinkabilityUnobservability-}
Pfitzman, A., Hansen, M.:
\newblock Anonymity, unlinkability, unobservability, pseudonymity, and identity
  management - a consolidated proposal for terminology.
\newblock http://dud.inf.tu-dresden.de/Anon\_Terminology.shtml (2006)

\bibitem{Barnum+Crepeau...-Authenticatio_of_q_mes:2002}
Barnum, H., Cr{\'e}peau, C., Gottesman, D., Smith, A., Tapp, A.:
\newblock Authentication of quantum messages.
\newblock In: FOCS 2002. (2002) quant-ph/0205128.

\bibitem{Ambainis+Buhrman...-Multi_quant_coin:2003}
Ambainis, A., Buhrman, H., Dodis, Y., Rohrig, H.:
\newblock Multiparty quantum coin flipping.
\newblock In: 19th IEEE Annual Conference on Computational Complexity (CCC'04).
  (2003)  250--259 quant-ph/0304112.

\bibitem{Bennett+Brassard...-Purif_noisy_entan:1995}
Bennett, C.H., Brassard, G., Popescu, S., Schumacher, B., Smolin, J.A.:
\newblock Purification of noisy entanglement and faithful teleportation via
  noisy channels.
\newblock Phys. Rev. Lett. \textbf{76} (1996)  722--725 quant-ph/9511027.

\bibitem{Ambainis.Gottesman-MinimumDistanceProblem-2003}
Ambainis, A., Gottesman, D.:
\newblock The minimum distance problem for two-way entanglement purification.
\newblock quant-ph/0310097, to appear in IEEE Transactions on Information
  Theory (2003)

\bibitem{Zhao.Chen.ea-Experimentaldemonstrationof-2004}
Zhao, Z., Chen, Y.A., Zhang, A.N., Yang, T., Briegel, H.J., Pan, J.W.:
\newblock Experimental demonstration of five-photon entanglement and
  open-destination teleportation.
\newblock Nature \textbf{430} (2004)  54--58

\end{thebibliography}
\appendix
\section{Proof of Lemma \ref{lem:qanon:disrupt_or_detect}}
\label{sec:Proof-Lemma}
In order to distinguish products of dual basis states corrupted participants have to perform some kind of measurement on $P_1, \dots,P_m$, which can be represented as a von Neumann measurement on some larger system (i.e. together with some ancillas) $P_1,\dots,P_m,A_1,\dots,A_k$, where ancillary systems are in some fixed pure state $\ket{a_1}$. It is easy to see that this measurement is the most general operation they can perform, since any unitary operation before measurement can be seen as a change of the measurement projection operators.

Let us suppose that they perform measurement using projectors $\{M_i\}_i=$ and let $\{\ket{b_j}\}_{j=1}^{2^{k+m}}$ be all eigenvectors of measurement projectors. We will express members of this basis as a superposition of vectors from product basis $\{\ket{+},\ket{-}\}^{\otimes m}\otimes\{\ket{a_i}\}_{i=1}^{2^k}$, where $\{\ket{a_i}\}_{i=1}^{2^k}$ is a basis on the ancillary system $A_1,\dots,A_k$. Let us express
\begin{equation}
\label{equ:anon_collapse_disruption}
\ket{b_j}=\sum_{i,l}\alpha_{i,l,j}\ket{\psi_l}\otimes\ket{a_i},
\end{equation}
where $\{\ket{\psi_l}\}_{l=1}^{2^m}$ is some enumeration of the basis states $\{\ket{+},\ket{-}\}^{\otimes m}$.
The fact that the measurement distinguishes products of dual basis states with good probability means that when there is a nonnegligible probability of obtaining projector $M\in\{M_i\}_i$ as an outcome, then this outcome must give answer about original state of the system $P_1,\dots,P_m$. It means $M$ restricted on systems $P_1,\dots,P_m$ is close to projector on some state $\ket{\psi_l}\in\{\ket{+},\ket{-}\}^{\otimes m}$. Otherwise $M$ has eigenvector $\ket{b_j}$ such that it has nonnegligible coefficient $\sum_i \alpha_{i,l',j}$ in Eq. \ref{equ:anon_collapse_disruption} corresponding to different basis states $\ket{\psi_{l'}}\ne\ket{\psi_l}$ and therefore $M$ might have been obtained even if the systems $P_1,\dots,P_m$ were in the state $\ket{\psi_{l'}}$.

This proves our statement since such a projector causes collapse of quantum systems $P_{m+1},\dots,P_n$ close to collapse caused by independent measurements in the dual basis with outcome $\ket{\psi_l}$.

\section{Proof of Theorem \ref{the:anon_EPR_generation}}
\label{sec:Proof-Theorem}
In order to prove sender's anonymity we will analyze all steps where his behavior differs from behavior of other potential senders. In Step $6\!:$ the sender does not measure state he received, but this is only local operation which cannot be detected. He measures all systems that were reported to be in one of the trap states. In Step $9\!:$ he reports all disagreements in states that were in advance reported to be trap states. He also announces optional disagreement in his measurement (although this implies that the distributor of the trap states was lying), but he measured trap states as all other honest potential senders and therefore his anonymity is not threatened.

In Step $11\!:$ the sender performs only local operations on his quantum systems what cannot be detected. In Step $14\!:$ the sender cooperates with the receiver on distillation of the shared EPR pairs. The sender is receiving classical communication through MTAR and performing local quantum operations and therefore his anonymity is not compromised. Analysis and discussion of Step $15\!:$ is in paragraphs following after Theorem \ref{the:anon_EPR_generation}.

To prove that each ensemble of EPR states originating from generalized GHZ state distributed by an honest participant is distillable we have to prove that the amount of noise disrupters can introduce into such an ensemble is below distillable threshold. In fact, we will show that at the cost of efficiency we can guarantee with arbitrarily high probability arbitrarily small amount\footnote{By amount of noise we mean the ratio between disrupted and pure EPR pairs in the ensemble.} of noise in all ensembles generated by honest participants.

Let $p$ be the probability that honest participant decides to operate in a trap mode, $(n-1)$ the number of rounds of Protocol \ref{alg:anonymous_EPR_generation}, $\theta$ the fraction of error repairable by the chosen entanglement distillation procedure (distillation succeeds if the number of errors in a group of $r$ systems will be lower that $\theta r$) and $\xi$ the probability that disrupter of the quantum system\footnote{We consider here maximal disruption, i.e. operation introducing maximal error in distillation procedure. Smaller error will allow better information about the trap state, but also causes smaller disruption, see Lemma \ref{lem:qanon:disrupt_or_detect}} guesses the trap state incorrectly.

Let us suppose that there is a collusion of disrupters disrupting generalized GHZ states distributed by a particular honest potential sender. In order to prevent distillation of this ensemble they have to disrupt $\theta m(1-p)$ systems (since on average only $(1-p)m$ rounds are in actual mode). When disrupting a particular system there is a probability $p\xi$ that disrupter will be detected. Probability that $\theta m(1-p)$ disruptions remain undetected is $$Prob(disr)=(1-p\xi)^{\theta m(1-p)},$$ i.e. the it decreases exponentially with $m$. Therefore we can choose arbitrary $Prob(disr)$ and $\theta$ and then choose sufficiently large $$m\geq\frac{\log_{1-p\xi}Prob(disr)}{\theta(1-p)}.$$

\end{document}